\documentclass[conference,10pt]{IEEEtran}
\IEEEoverridecommandlockouts
\usepackage{cite}
\usepackage{amsmath,amssymb,amsfonts}
\usepackage{algorithmic}
\usepackage{graphicx}
\usepackage{textcomp}
\usepackage{xcolor}
\usepackage{authblk}

\usepackage{amsthm}

\usepackage{booktabs}

\renewcommand{\v}[1]{{\bf #1}}
\newcommand{\m}[1]{{\bf #1}}
\newcommand{\R}{\mathbb{R}}

\renewcommand{\O}{\mathcal{O}}
\DeclareMathOperator{\tr}{Tr}

\usepackage{algorithm}
\usepackage{algorithmic}

\usepackage{etex,etoolbox}
\usepackage{thmtools}
\usepackage{environ}

\makeatletter
\providecommand{\@fourthoffour}[4]{#4}
\newcommand\fixstatement[2][\proofname\space of]{%
  \ifcsname thmt@original@#2\endcsname
    \AtEndEnvironment{#2}{%
      \xdef\pat@label{\expandafter\expandafter\expandafter
        \@fourthoffour\csname thmt@original@#2\endcsname\space\@currentlabel}%
      \xdef\pat@proofof{\@nameuse{pat@proofof@#2}}%
    }%
  \else
    \AtEndEnvironment{#2}{%
      \xdef\pat@label{\expandafter\expandafter\expandafter
        \@fourthoffour\csname #1\endcsname\space\@currentlabel}%
      \xdef\pat@proofof{\@nameuse{pat@proofof@#2}}%
    }%
  \fi
  \@namedef{pat@proofof@#2}{#1}%
}

\globtoksblk\prooftoks{1000}
\newcounter{proofcount}

\NewEnviron{proofatend}{%
  \edef\next{%
    \noexpand\begin{proof}[\pat@proofof\space\pat@label]%
    \unexpanded\expandafter{\BODY}}%
  \global\toks\numexpr\prooftoks+\value{proofcount}\relax=\expandafter{\next\end{proof}}
  \stepcounter{proofcount}}

\def\printproofs{%
  \count@=\z@
  \loop
    \the\toks\numexpr\prooftoks+\count@\relax
    \ifnum\count@<\value{proofcount}%
    \advance\count@\@ne
  \repeat}
\makeatother

\declaretheorem[style=plain,name=Theorem,]{thm}
\newtheorem{lem}{Lemma}
\fixstatement{thm}
\fixstatement[Demonstration of]{lem}

\newcommand{\fp}{_\textup{FP}}

\def\BibTeX{{\rm B\kern-.05em{\sc i\kern-.025em b}\kern-.08em
    T\kern-.1667em\lower.7ex\hbox{E}\kern-.125emX}}
\begin{document}

\title{Biologically Plausible Online Principal Component Analysis Without Recurrent Neural Dynamics
}

\author[1]{Victor Minden}
\author[1]{Cengiz Pehlevan}
\author[*1,2]{Dmitri B. Chklovskii}

\affil[1]{Center for Computational Biology, Flatiron Institute, New York, NY 10010 \authorcr Email: {\tt\small \{vminden, cpehlevan, mitya\}@flatironinstitute.org}\vspace{1.5ex}}
\affil[2]{Neuroscience Institute, NYU Langone Medical Center, New York, NY 10016  \vspace{-2ex}}


\maketitle

\begin{abstract}
Artificial neural networks that learn to perform Principal Component Analysis (PCA) and related tasks using strictly local learning rules have been previously derived based on the principle of similarity matching: similar pairs of inputs should map to similar pairs of outputs.  However, the operation of these networks (and of similar networks) requires a fixed-point iteration to determine the output corresponding to a given input, which means that dynamics must operate on a faster time scale than the variation of the input.  Further, during these fast dynamics such networks typically ``disable'' learning, updating synaptic weights only once the fixed-point iteration has been resolved.  Here, we derive a network for PCA-based dimensionality reduction that avoids this fast fixed-point iteration.  The key novelty of our approach is a modification of the similarity matching objective to encourage near-diagonality of a synaptic weight matrix.  We then approximately invert this matrix using a Taylor series approximation, replacing the previous fast iterations.  In the offline setting, our algorithm corresponds to a dynamical system, the stability of which we rigorously analyze.  In the online setting (i.e., with stochastic gradients), we map our algorithm to a familiar neural network architecture and give numerical results showing that our method converges at a competitive rate.  The computational complexity per iteration of our online algorithm is linear in the total degrees of freedom, which is in some sense optimal.
\end{abstract}


\section{Introduction}
Our brain effortlessly processes high-dimensional data streamed to it by sensory organs to make behaviorally relevant decisions. Yet, an algorithmic model of online neural computation that satisfies known biological constraints does not exist. Recently, a first step towards such a model has been made by deriving neural networks, including both neuronal activity dynamics and synaptic learning rules, from the similarity matching objective \cite{derivation,offdiag,whydo}.  In particular, single-layer neural networks for linear dimensionality reduction have, for the first time, been derived from a principled objective in a manner leading to local learning rules (where a synaptic weight is updated based only on the activity of only the two neurons that synapse connects).

However, the operation of existing similarity matching networks, as well as other neural network algorithms for dimensionality reduction \cite{linsker1992,ica}, is based on an iterative update of neuronal activity or synaptic weights via recurrent dynamics which must converge by the time the next stimulus arrives. For fast senses, such as audition and vision, there is no clear separation in time scales between the correlation time of the input signal and the neural dynamics to satisfy such requirement. Therefore, deriving a dimensionality reduction network with iteration-free dynamics is an important problem.

In this paper, we address this problem by deriving a biologically plausible network from a novel objective function. The iterative dynamics in similarity matching networks stem from a biologically plausible iterative implementation of matrix inversion (i.e., solving a linear system).  By modifying the similarity matching objective to eliminate degeneracies of the solution with respect to rotations and reflections, we identify a particular family of solutions for which the matrix to be inverted is, at a fixed point, diagonal, which motivates our modified algorithm that substitutes iterative dynamics with a low-order expansion of the matrix inverse.

While removing degeneracies from neural networks for similarity matching based on modifying the objective has been proposed previously \cite{offdiag}, such networks still required iterative dynamics to compute the output for each input.  Similar symmetry-breaking techniques have been used for other PCA networks not based on similarity matching, and we note in particular the Weighted Subspace Criterion of Oja, Ogawa, and Wangviwattana \cite{ojaogawa} and work by Xu \cite{xu}.

There exists a wealth of relevant literature on neural approaches to online PCA and related tasks.  We direct the reader to the review by Qiu et al.\ \cite{qiu} for a full treatment of such algorithms, and mention here those that are most directly relevant. These include work by Oja \cite{ojaold,oja1989,oja1992}, the APEX algorithm \cite{apex}, Sanger's rule \cite{gha}, F\"oldi\'ak's network \cite{foldiak}, and work by Linsker \cite{linsker2005improved}.  Compared to these, only our approach starts from a principled objective function and arrives at both local (anti-)Hebbian learning rules and a neural network architecture for multi-component PCA without requiring fast recurrent neural dynamics.

\section{Problem Formulation}\label{sec:prob}
In the following, bold lower-case variables refer to column vectors, e.g., $\v{x}_t\in\R^N$, whereas bold upper-case variables refer to matrices, e.g., $\m{X}\in\R^{N\times T}.$  We use $\|\cdot\|$ for the standard Euclidean vector norm as well as for the matrix Frobenius norm, i.e., $\|\v{x}_t\|^2 = \v{x}_t^\top\v{x}_t$ and $\|\m{X}\|^2 = \tr(\m{X}^\top\m{X})$.

\subsection{Optimization Problems}
Given an input data matrix $\m{X}\in\R^{N\times T}$ with columns $\{\v{x}_t\}_{t=1}^T\subset\R^N$, the Principal Subspace Projection (PSP) problem is to find a low-dimensional embedding $\m{Y}\in\R^{K\times T}$ of $\m{X}$ that optimally preserves inner products between pairs of input vectors, i.e., solve
\begin{equation}\label{eq:pca}
\min_{\m{Y}}  \|\m{X}^\top\m{X} - \m{Y}^\top\m{Y}\|^2 = \min_{\{\v{y}_t\}_{t=1}^T}  \sum_{t,t'} \left(\v{x}_t^\top\v{x}_{t'} - \v{y}_t^\top\v{y}_{t'}\right)^2.
\end{equation}
By expanding the norm and dropping terms independent of $\m{Y}$ we obtain the equivalent optimization problem
\begin{equation}\label{eq:pspobj}
\min_{\m{Y}}  -2\tr\left(\m{X}^\top\m{X}\m{Y}^\top\m{Y}\right) + \tr\left(\m{Y}^\top\m{Y}\m{Y}^\top\m{Y}\right).
\end{equation}
It is well known that solutions to \eqref{eq:pca} and thus \eqref{eq:pspobj} are given by PCA, i.e., if $\m{X} = \m{U}\m{\Sigma}\m{V}^\top$ is a Singular Value Decomposition (SVD) then
$\m{Y}_\textup{opt}=\m{Q}\m{U}_K^\top\m{X}$, where $\m{Q}\in\R^{K\times K}$ is an arbitrary orthogonal matrix and $\m{U}_K\in\R^{T\times K}$ contains the first $K$ columns of $\m{U}$ with corresponding singular values $\{\sigma_k\}_{k=1}^K.$

To break the degeneracy of \eqref{eq:pspobj}, we introduce a diagonal matrix $\m{\Lambda}\in\R^{K\times K}$ with distinct positive diagonal entries $\lambda_1>\lambda_2>\dots>\lambda_K$ and obtain a related problem
\begin{equation}\label{eq:pspobjlambda}
\min_{\m{Y}}  -2\tr\left(\m{X}^\top\m{X}\m{Y}^\top\m{Y}\right) + \tr\left(\m{Y}^\top\m{\Lambda}^{-1}\m{Y}\m{Y}^\top\m{\Lambda}^{-1}\m{Y}\right).
\end{equation}
Compared to \eqref{eq:pspobj}, the solution to \eqref{eq:pspobjlambda} is unique up to sign ambiguity under certain assumptions on $\m{X}$.

\begin{lem}\label{lem:uniquepsp}
  Suppose that the top $K$ singular values of $\m{X}$ are unique, i.e., $\sigma_1(\m{X}) > \sigma_2(\m{X}) > \dots > \sigma_{K+1}(\m{X}).$ Then the optimal solutions to \eqref{eq:pspobjlambda} are given by
  \begin{equation*}
    \m{Y}_\textup{opt}= \m{\Lambda}\m{S}\m{U}_K^\top\m{X},
  \end{equation*}
  where $\m{S}\in\R^{K\times K}$ is any diagonal matrix with diagonal entries in $\{1,-1\}$.
\end{lem}
\begin{proofatend}
  Taking a derivative in $\m{Y}$, we obtain the first-order stationarity condition
  \begin{equation*}
    -4\m{Y}\m{X}^\top\m{X} + 4\m{\Lambda}^{-1}\m{Y}\m{Y}^\top\m{\Lambda}^{-1}\m{Y} = \v{0}.
  \end{equation*}
  Right-multiplying both sides of this equation by $\m{Y}^\top\m{\Lambda}^{-1}$ and rearranging gives
  \begin{equation*}
  \m{Y}\m{X}^\top\m{X}\m{Y}^\top\m{\Lambda}^{-1} = \m{\Lambda}^{-1}\m{Y}\m{Y}^\top\m{\Lambda}^{-1}\m{Y}\m{Y}^\top\m{\Lambda}^{-1},
\end{equation*}
  where we observe that the left-hand side must be symmetric.  We conclude that $\m{\Lambda}^{-1}$ and $\m{Y}\m{X}^\top\m{X}\m{Y}^\top$ commute and thus share the same eigenvectors.  Therefore, $\m{Y}^\top$ diagonalizes $\m{X}^\top\m{X}$, i.e., each nonzero row of $\m{Y}$ is an eigenvector of $\m{X}^\top\m{X}$ and the rows of $\m{Y}$ are pairwise orthogonal. With $\m{X}=\m{U}\m{\Sigma}\m{V}^\top$ as the SVD of $\m{X}$, we write $\m{Y}\m{Y}^\top=\m{D}$ and $\m{Y}\m{X}^\top\m{X}\m{Y}^\top=\m{D}\m{\widetilde\Sigma}^2_K$, where $\m{D}$ is a diagonal matrix with nonnegative diagonal entries and $\m{\widetilde\Sigma}_K$ is a $K\times K$ principal submatrix of $\m{\Sigma}$.  Plugging this representation into \eqref{eq:pspobjlambda}, we obtain
  \begin{equation*}
    -2\tr\left(\m{D}\m{\widetilde\Sigma}^2_K\right) + \tr\left(\m{D}^2\m{\Lambda}^{-2}\right),
  \end{equation*}
  which is minimized by choosing $\m{\widetilde\Sigma}_K = \m{\Sigma}_K$ and $\m{D}=\m{\Lambda}^2\m{\Sigma}^2_K$.  The result follows.
\end{proofatend}

Using \eqref{eq:pspobjlambda} as a starting point, we obtain a mixed min-max objective from \eqref{eq:pspobjlambda} by following the similarity matching framework \cite{whydo}, introducing two auxiliary variables $\m{M}\in\R^{K\times K}$ and $\m{W}\in\R^{K\times N}$ to obtain
\begin{equation}\label{eq:pspobjlambda2}
  \begin{aligned}
\min_{\m{Y},\m{W}} \max_{\m{M}}\;  2&\tr\left({\m{W}^\top\m{W}}\right)-4\tr\left(\m{X}^\top\m{W}^\top\m{Y}\right)\\\; + 2&\tr\left(\m{Y}^\top\m{M}\m{Y}\right) - \tr\left(\m{\Lambda}\m{M}^\top\m{\Lambda}\m{M}\right).
\end{aligned}
\end{equation}
To see that \eqref{eq:pspobjlambda2} is equivalent to \eqref{eq:pspobjlambda} note that the subproblems in $\m{W}$ and $\m{M}$ have unique solutions $\m{W}_\textup{opt}(\m{Y}) =\m{Y}\m{X}^\top$ and $\m{M}_\textup{opt}(\m{Y})=\m{\Lambda}^{-1}\m{Y}\m{Y}^\top\m{\Lambda}^{-1}$, which follow from first-order optimality conditions.

As a related problem to PSP, we consider also Principal Subspace Whitening (PSW),
\begin{equation*}
\min_{\m{Y}}  \|\m{X}^\top\m{X} - \m{Y}^\top\m{Y}\|^2 \quad \text{s.t.} \quad \m{Y}\m{Y}^\top = \m{I},
\end{equation*}
which adds to PSP the constraint that the output features must be whitened, i.e., rows of $\m{Y}$ must be orthogonal.  Solutions to PSW are given again in terms of PCA with an arbitrary orthogonal transform $\m{Q}$, though now $\m{Y}_\textup{opt}=\m{Q}\m{\Sigma}_K^{-1}\m{U}_K^\top\m{X}$ to ensure whitened output.  To break the degeneracy we replace the constraint with $\m{Y}\m{Y}^\top = \m{\Lambda}^2$ with diagonal matrix $\m{\Lambda}$ as previously described, leading to
\begin{equation}\label{eq:pswobjlambda}
\min_{\m{Y}}  \|\m{X}^\top\m{X} - \m{Y}^\top\m{Y}\|^2 \quad \text{s.t.} \quad \m{Y}\m{Y}^\top = \m{\Lambda}^2.
\end{equation}
We enforce the constraint using a Lagrange multiplier, $\bf M$, and, following Pehlevan, Sengupta, and Chklovskii \cite{whydo}, we ultimately obtain the min-max objective
\begin{equation}\label{eq:pswobjlambda2}
  \begin{aligned}
\min_{\m{Y},\m{W}} \max_{\m{M}}\;  &\tr\left({\m{W}^\top\m{W}}\right)-2\tr\left(\m{X}^\top\m{W}^\top\m{Y}\right)\\\; + &\tr\left(\m{M}\left[\m{Y}\m{Y}^\top - \m{\Lambda}^2\right]\right).
\end{aligned}
\end{equation}

\begin{lem}\label{lem:uniquepsw}
  Suppose the same setting as Lemma \ref{lem:uniquepsp}. Then optimal solutions to \eqref{eq:pswobjlambda} are given by
  \begin{equation*}
    \m{Y}_\textup{opt}= \m{\Lambda}\m{S}\m{\Sigma}_K^{-1}\m{U}_K^\top\m{X},
  \end{equation*}
  where $\m{S}\in\R^{K\times K}$ is any diagonal matrix with diagonal entries in $\{1,-1\}$.
\end{lem}
\begin{proofatend}
The result follows from writing an SVD of $\m{Y}$ and then first, observing that $\m{Y}\m{Y}^\top = \m{\Lambda}^2$ implies that left singular vectors of $\m{Y}$ are coordinate vectors and singular values of $\m{Y}$ are on the diagonal of $\m{\Lambda}$ and second, observing the optimal choice of right singular vectors is given by setting them to corresponding right singular vectors of $\m{X}$ (this is the two-sided orthogonal Procrustes problem \cite{schonemann}).
\end{proofatend}

We emphasize that by introducing the diagonal matrix $\m{\Lambda}$ to the original PSP and PSW problems the degeneracy of the solution with respect to rotation is eliminated and we are left with only a sign ambiguity for the vector corresponding to each principal component.

\subsection{Dynamical Systems}
To derive an optimization scheme for PSP we use the fact that \eqref{eq:pspobjlambda2} exhibits strong duality in $\m{Y}$ and $\m{M}$ such that minimization over $\m{Y}$ may be performed prior to maximization over $\m{M}$ \cite[Proposition 1]{whydo}.  The first-order optimality conditions for $\m{Y}$ then give\footnote{$\m{M}$ in \eqref{eq:pspobjlambda2} can be constrained to be invertible, as discussed in Pehlevan et al.\cite{whydo}.} $\m{Y}_\textup{opt}(\m{M},\m{W}) = \m{M}^{-1}\m{W}\m{X}$.  Plugging this back into \eqref{eq:pspobjlambda2} yields a saddle-point problem
\begin{equation}\label{eq:saddle}
  \min_{\m{W}} \max_{\m{M}}\; \textup{L}(\m{M},\m{W}),
\end{equation}
with objective
\begin{align*}
 \textup{L}(\m{M},\m{W}) &=\;  2\tr\left({\m{W}^\top\m{W}}\right) - \tr\left(\m{\Lambda}\m{M}^\top\m{\Lambda}\m{M}\right)\\
&\phantom{=}  -2\tr\left(\m{X}^\top\m{W}^\top\m{M}^{-1}\m{W}\m{X}\right)\\
 &= 2\tr\left({\m{W}^\top\m{W}}\right) - \tr\left(\m{\Lambda}\m{M}^\top\m{\Lambda}\m{M}\right)\\
&\phantom{=}  -2\tr\left(\m{X}^\top\m{W}^\top\m{Y}_\textup{opt}(\m{M},\m{W})\right).
\end{align*}
From \eqref{eq:saddle} we can obtain an offline dynamical system by prescribing gradient ascent dynamics to $\m{M}$ and gradient descent dynamics to $\m{W}$, i.e.,
\begin{equation}\label{eq:dyn1psp}
  \begin{array}{r@{}l}
  \m{Y}(s) &{}= \m{M}^{-1}(s)\m{W}(s)\m{X},\\[3pt]
  \frac{d\m{W}(s)}{ds} &{}= \m{Y}(s)\m{X}^\top - \m{W}(s),\\[3pt]
  \tau\frac{d\m{M}(s)}{ds} &{}= \m{Y}(s)\m{Y}(s)^\top - \m{\Lambda}\m{M}(s)\m{\Lambda},
\end{array}
\end{equation}
where $\tau>0$ is a constant setting the time scale for dynamics of $\m{M}$ relative to that of dynamics of $\m{W}$.
By construction, fixed points of \eqref{eq:dyn1psp} are critical points of \eqref{eq:saddle}.

Applying the same basic framework as above to the min-max objective for PSW in \eqref{eq:pswobjlambda2}, we get a similar set of offline dynamics for the whitening task:
\begin{equation}\label{eq:dyn1psw}
  \begin{array}{r@{}l}
  \m{Y}(s) &{}= \m{M}^{-1}(s)\m{W}(s)\m{X},\\[3pt]
  \frac{d\m{W}(s)}{ds} &{}= \m{Y}(s)\m{X}^\top - \m{W}(s),\\[3pt]
  \tau\frac{d\m{M}(s)}{ds} &{}= \m{Y}(s)\m{Y}(s)^\top - \m{\Lambda}^2.
\end{array}
\end{equation}

\section{Results}
To obtain neural networks with biologically plausible learning rules for PCA, we first derive some properties of the dynamical systems \eqref{eq:dyn1psp} and \eqref{eq:dyn1psw} and use insights from those to obtain related dynamical systems that can be then converted to online algorithms.  We defer proofs to the appendix.

\subsection{Stability of Dynamical Systems}
If we take $\m{\Lambda}=\m{I}$, the dynamical system \eqref{eq:dyn1psp} reduces to that of Pehlevan, Sengupta, and Chklovskii~\cite[Theorem 1]{whydo}. They proved that, at any stable fixed point $(\m{M}\fp,\m{W}\fp)$, $\m{M}\fp^{-1}\m{W}\fp$ has orthogonal rows spanning the principal subspace of $\m{X}$.  For $\m{\Lambda}$ with distinct diagonal entries, a stronger result holds.

\begin{lem}\label{lem:psp1}
  Let $(\m{M}\fp,\m{W}\fp)$ be a fixed point of the dynamical system \eqref{eq:dyn1psp} and define $\m{F}\fp \equiv \m{M}\fp^{-1}\m{W}\fp$. Then the matrix $\m{M}\fp$ is diagonal with eigenvalues of $\m{C} \equiv \m{X}\m{X}^\top$ on the diagonal, and the corresponding eigenvectors form the rows of the matrix $\m{F}\fp$ after scaling them so that $\m{F}\fp\m{F}\fp^\top = \m{\Lambda}^2$.
\end{lem}
\begin{proofatend}
   At a fixed point, we have from \eqref{eq:dyn1psp} that $\m{F}\fp\m{C} = \m{W}\fp$ and
   $\m{F}\fp\m{C}\m{F}\fp^\top = \m{\Lambda}\m{M}\fp\m{\Lambda}.$
   Together these imply $\m{W}\fp\m{F}\fp^\top = \m{\Lambda}\m{M}\fp\m{\Lambda}$, but by definition $\m{F}\fp \equiv \m{M}\fp^{-1}\m{W}\fp$ so this gives $\m{W}\fp\m{W}\fp^\top\m{M}\fp^{-1} = \m{\Lambda}\m{M}\fp\m{\Lambda}$ and thus $\m{\Lambda}\m{M}\fp\m{\Lambda}\m{M}\fp = \m{M}\fp\m{\Lambda}\m{M}\fp\m{\Lambda}$.  Therefore, $\m{M}\fp\m{\Lambda}\m{M}\fp$ and $\m{\Lambda}$ commute, from which it follows that $\m{M}\fp\m{\Lambda}\m{M}\fp$ is diagonal.  Then, we observe
   $$
   \left[\left(\m{M}\fp\m{\Lambda}\m{M}\fp\right)\m{\Lambda}\right]\m{M}\fp = \m{M}\fp\left[\m{\Lambda}\left(\m{M}\fp\m{\Lambda}\m{M}\fp\right)\right],
   $$
   which shows $\m{M}\fp$ commutes with the bracketed diagonal matrix and is thus itself diagonal.  The results for $\m{F}\fp$ then follow from $\m{F}\fp\m{C}=\m{W}\fp=\m{M}\fp\m{F}\fp$ and $\m{F}\fp\m{C}\m{F}\fp^\top = \m{\Lambda}\m{M}\fp\m{\Lambda}$.
\end{proofatend}

Considering small perturbations around a fixed point, i.e., $(\m{M}\fp+\delta\m{M},\m{W}\fp+\delta\m{W})$, a formal linear stability analysis of \eqref{eq:dyn1psp} gives the following result.
\begin{thm}\label{thm:psp1}
  Consider the same setting as Lemma \ref{lem:uniquepsp}.  Then, for sufficiently small $\tau$, stable fixed points of \eqref{eq:dyn1psp} are $(\m{M}\fp,\m{W}\fp)$ such that the rows of $\m{F}_\textup{FP}$ span the principal subspace of $\m{X}$.  In particular, we have a stable fixed point when $\m{F}_\textup{FP}= \m{\Lambda}\m{S}\m{U}_K^\top$ (as in Lemma \ref{lem:uniquepsp}), i.e., each row of $\m{F}\fp$ is a (signed) multiple of the corresponding singular vector of $\m{X}$.
\end{thm}
\begin{proofatend}
  First we show that a necessary condition for stability is that the rows of $\m{F}\fp$ span the principal $k$-dimensional subspace of $\m{C}$.  The linearization of \eqref{eq:dyn1psp} is
  \begin{equation}\label{eq:deltadyn}
    \begin{array}{r@{}l}
  \frac{d(\delta\m{W})}{ds} &{}= (\delta\m{F})\m{C} - (\delta\m{W}),\\[3pt]
  \tau\frac{d(\delta\m{M})}{ds} &{}=(\delta\m{F})\m{C}\m{F}\fp^{\top} + \m{F}\fp\m{C}(\delta\m{F})^\top - \m{\Lambda}(\delta \m{M})\m{\Lambda},
  \end{array}
  \end{equation}
  where $\delta\m{F} \equiv \m{M}\fp^{-1}(\delta\m{W}) - \m{M}\fp^{-1}(\delta\m{M})\m{F}\fp$ and we have dropped explicit dependencies on $s$.  With some algebra, we find
  \begin{equation}\label{eq:deltafdyn}
    \begin{aligned}
  \frac{d(\delta\m{F})}{ds} = &\frac{\m{M}\fp^{-1}}{\tau}\left[\m{\Lambda}(\delta \m{M})\m{\Lambda}  - (\delta\m{F})\m{C}\m{F}\fp^\top -\m{F}\fp\m{C}(\delta\m{F})^\top\right]\m{F}\fp\\
  &+\m{M}\fp^{-1}(\delta\m{F})\m{C} - \delta \m{F} - \m{M}\fp^{-1}(\delta\m{M})\m{F}\fp.
\end{aligned}
\end{equation}
We decompose $\delta\m{F} = (\delta\m{N})\m{F}\fp + (\delta\m{B})\m{G}\fp$, where the rows of the matrix $\m{G}\fp~\in~\R^{(N-K)\times N}$ are $N-K$ orthonormal eigenvectors of $\m{C}$ that are orthogonal to rows of $\m{F}\fp$.  Right-multiplying \eqref{eq:deltafdyn} by $\m{G}\fp^\top$ gives
\begin{equation}\label{eq:bdyn}
  \frac{d(\delta\m{B})}{ds} =\m{M}\fp^{-1}(\delta\m{B})\m{D} - \delta \m{B},
\end{equation}
where $\m{D}$ is a diagonal matrix with the eigenvalues corresponding to $\m{G}\fp$, i.e., $\m{G}\fp\m{C}\m{G }\fp^\top = \m{D}.$  A stability analyis of $\eqref{eq:bdyn}$ shows stability requires $\lambda_\text{min}(\m{M}\fp) > \lambda_\text{max}(\m{D})$, which implies the diagonal of $\m{M}$ contains the top $K$ eigenvalues of $\m{C}$.

Now we show that, for sufficiently small $\tau$, points with $\m{F}\fp$ of the form $\m{F}\fp= \m{\Lambda}\m{S}\m{U}_K^\top$ are stable fixed points by analyzing the perturbation $\delta{\m{N}}$.  Using the fact that $\m{F}\fp\m{C} = \m{M}\fp\m{F}\fp$ and right-multiplying by $\m{F}\fp^\top\m{\Lambda}^{-2}$ we obtain from \eqref{eq:deltafdyn}
\begin{equation}\label{eq:deltaNdyn}
  \begin{aligned}
  \frac{d(\delta\m{N})}{ds} = &\frac{\m{M}\fp^{-1}}{\tau}\left[\m{\Lambda}(\delta \m{M})\m{\Lambda}  - (\delta\m{N})\m{\Lambda}^2\m{M}\fp -\m{M}\fp\m{\Lambda}^2(\delta\m{N})^\top\right]\\
&+\m{M}\fp^{-1}(\delta\m{N})\m{M}\fp - \delta \m{N} - \m{M}\fp^{-1}(\delta\m{M})
\end{aligned}
\end{equation}
and from \eqref{eq:deltadyn}
\begin{equation}\label{eq:deltamdyn}
\tau\frac{d(\delta\m{M})}{ds} =(\delta\m{N})\m{\Lambda}^2\m{M}\fp + \m{M}\fp\m{\Lambda}^2(\delta\m{N})^\top - \m{\Lambda}(\delta \m{M})\m{\Lambda},
\end{equation}
which together form the dynamical system we can use to analyze stability of $\delta\m{N}$ perturbations.  Because $\m{M}\fp$ is diagonal, we see that the system consists of small $3\times 3$ subsystems describing the coupled dynamics of $(\delta\m{M}_{ij},\delta\m{N}_{ij}, \delta\m{N}_{ji})$, where individual element dynamics in this subsystem are
\begin{equation*}
  \begin{aligned}
    \tau\frac{d(\delta\m{M}_{ij})}{ds} &= \lambda_{j}^2 m_j(\delta\m{N}_{ij}) + m_i\lambda_i^2(\delta\m{N}_{ji}) - \lambda_i\lambda_j(\delta \m{M}_{ij}),\\
    \frac{d(\delta\m{N})_{ij}}{ds} = &\frac{1}{m_{i}\tau}\left[\lambda_i(\delta \m{M}_{ij})\lambda_j  - (\delta\m{N}_{ij})\lambda_j^2 m_j -m_i\lambda_i^2(\delta\m{N}_{ji})\right]\\
  &+\frac{1}{m_i}(\delta\m{N}_{ij})m_j - \delta \m{N}_{ij} - \frac{1}{m_i}(\delta\m{M}_{ij}),\\
  \frac{d(\delta\m{N})_{ji}}{ds} = &\frac{1}{m_{j}\tau}\left[\lambda_j(\delta \m{M}_{ji})\lambda_i  - (\delta\m{N}_{ji})\lambda_i^2 m_i -m_j\lambda_j^2(\delta\m{N}_{ij})\right]\\
&+\frac{1}{m_j}(\delta\m{N}_{ji})m_i - \delta \m{N}_{ji} - \frac{1}{m_j}(\delta\m{M}_{ji}).
  \end{aligned}
\end{equation*}
As an aside, we note that we need only consider symmetric $\delta\m{M}$, since $\m{M}$ is constrained throughout to be symmetric.
Denoting by $\m{J}\fp^{(ij)}$ the Jacobian corresponding to the subsystem involving $\delta\m{M}_{ij}$, we have
\begin{equation*}
  \m{J}\fp^{(ij)} \equiv \left(\begin{array}{ccc} -\frac{\lambda_i\lambda_j}{\tau}&\frac{\lambda_j^2m_j}{\tau}&\frac{\lambda_i^2m_i}{\tau}\\
  \frac{\lambda_i\lambda_j}{m_i\tau} - \frac{1}{m_i}&\frac{m_j}{m_i}-\frac{m_j\lambda_j^2}{m_i\tau}-1& -\frac{\lambda_i^2}{\tau}\\
\frac{\lambda_i\lambda_j}{m_j\tau} - \frac{1}{m_j}&-\frac{\lambda_j^2}{\tau}&\frac{m_i}{m_j}-\frac{m_i\lambda_i^2}{m_j\tau}-1\end{array}\right).
\end{equation*}
We are interested in the eigenvalues of $\m{J}\fp^{(ij)}$ but it is easier to consider the eigenvalues of the similar matrix
\begin{equation*}
  \m{H}\fp^{(ij)}\equiv
  \left(
  \begin{array}{ccc}
    -\frac{\lambda_i\lambda_j}{\tau} -\frac{\lambda_j^2m_j}{m_i\tau} - \frac{\lambda_i^2m_i}{m_j\tau}&\frac{\lambda_j^2m_j}{m_i\tau}&\frac{\lambda_i^2m_i}{m_j\tau}\\
 - \frac{m_j}{m_i}&\frac{m_j}{m_i}-1& 0\\
 - \frac{m_i}{m_j}&0&\frac{m_i}{m_j}-1
 \end{array}\right),
\end{equation*}
where the similarity transform is $\m{H}\fp^{(ij)}= \m{S}\m{J}\fp^{(ij)}\m{S}^{-1}$ with
\begin{equation*}
  \m{S} \equiv   \left(
    \begin{array}{ccc} 1&0 &0 \\ 1&m_i &0 \\ 1& 0&m_j \end{array}\right).
\end{equation*}


Letting $t_{ij} = m_i/m_j$, we find that the characteristic polynomial $\chi(s)$ of $\m{H}\fp^{(ij)}$ is given by
\begin{align*}
  \chi(s) = - s^3 + b s^2 + c s + d,
\end{align*}
with
\begin{align*}
  b &= -(t_{ij}-1)(t_{ji}-1) -\frac{[\lambda_i\lambda_j+\lambda_j^2t_{ji} +\lambda_i^2t_{ij}]}{\tau},\\
  d &= -\frac{[\lambda_j^2(t_{ji}-1) +\lambda_i^2(t_{ij}-1)+\lambda_i\lambda_j(t_{ij}-1)(t_{ji}-1)]
}{\tau},\\
  c &= b+d + \frac{\lambda_i\lambda_j}{\tau}.
\end{align*}
For sufficiently small $\tau$, we see that $b$ is negative and we can show that $d$ is negative as long as $t_{ij}>t_{ji}$ implies $\lambda_i > \lambda_j$ (which holds only when the ordering implied by $\m{\Lambda}$ is matched by that of $\m{M}\fp$).  By inspection, in this case $c$ is also negative, which means by Descartes' rule of signs we have no positive real roots.  Further, it can be verified that the polynomial discriminant of $\chi(s)$ is positive (and thus we have three real roots) as long as $b+d<0$.  We conclude that for sufficiently small $\tau$ we have three real negative roots and thus a stable fixed point.
\end{proofatend}

Analogously to PSP, if we take $\m{\Lambda}=\m{I}$, the PSW dynamical system \eqref{eq:dyn1psw} reduces to that of Pehlevan, Sengupta, and Chklovskii~\cite[Theorem 2]{whydo}. Similarly, they proved that, at any stable fixed point $(\m{M}\fp,\m{W}\fp)$, $\m{M}\fp^{-1}\m{W}\fp$ has orthogonal rows spanning the principal subspace of $\m{X}$.  For $\m{\Lambda}$ with distinct diagonal entries, a stronger result again holds and we can perform a similar formal linear stability analysis.
\begin{lem}\label{lem:psw1}
  Let $(\m{M}\fp,\m{W}\fp)$ be a fixed point of the dynamical system \eqref{eq:dyn1psw} and define $\m{F}\fp \equiv \m{M}\fp^{-1}\m{W}\fp$. Then the matrix $\m{M}\fp$ is diagonal with eigenvalues of $\m{C} \equiv \m{X}\m{X}^\top$ on the diagonal, and the corresponding eigenvectors form the rows of the matrix $\m{F}\fp$ after scaling them so that $\m{F}\fp\m{C}\m{F}\fp^\top = \m{\Lambda}^2$.
\end{lem}
\begin{proofatend}
  At a fixed point, we have from \eqref{eq:dyn1psw} that $\m{F}\fp\m{C} = \m{W}\fp$ and $\m{F}\fp\m{C}\m{F}\fp^\top = \m{\Lambda}^2$.  Since $\m{F}\fp = \m{M}^{-1}\fp\m{W}\fp,$ this implies $\m{M}\fp\m{F}\fp\m{F}^\top\fp = \m{\Lambda}^2$, which we rearrange to conclude that $\m{M}\fp$ and $\m{\Lambda}$ commute and so $\m{M}\fp$ is diagonal and the rows of $\m{F}\fp$ are orthogonal and scaled such that $\m{F}\fp\m{F}^\top\fp = \m{M}\fp^{-1}\m{\Lambda}^2$.
\end{proofatend}

\begin{thm}\label{thm:psw1}
  Consider the same setting as Lemma \ref{lem:uniquepsp}.  Then for sufficiently small $\tau$, stable fixed points of \eqref{eq:dyn1psw} are $(\m{M}\fp,\m{W}\fp)$ such that the rows of $\m{F}_\textup{FP}$ span the principal subspace of $\m{X}$.  In particular, we have a stable fixed point when $\m{F}_\textup{FP}= \m{\Lambda}\m{S}\m{\Sigma}_K^{-1}\m{U}_K^\top$ (as in Lemma \ref{lem:uniquepsw}), i.e., each row of $\m{F}\fp$ is a (signed) multiple of the corresponding singular vector of $\m{X}$.
\end{thm}
\begin{proofatend}
This proof mirrors the proof of Theorem \ref{thm:psp1}, up to the point where we must consider the eigenvalues of the Jacobian of the linearized system.  At that point, direct computation reveals the charactistic polynomial
\begin{align*}
  \chi(s) = - s^3 + b s^2 + c s + d
\end{align*}
with
\begin{align*}
  b &= (t_{ij}-1 + t_{ji}-1) - \frac{m_i^{-1}\lambda_j^2 + m_j^{-1}\lambda_i^2}{\tau},\\
  d &= \frac{(m_i^{-1}-m_j^{-1})(\lambda_i^2 - \lambda_j^2)}{\tau},\\
  c &= b+d,
\end{align*}
where variables are as defined in the proof of Theorem \ref{thm:psp1}, and the proof concludes similarly.
\end{proofatend}

\subsection{Avoiding Matrix Inversion}

The dynamics \eqref{eq:dyn1psp} and \eqref{eq:dyn1psw} seem to require the matrix inverse $\m{M}^{-1}(s)$ which is cumbersome for a biologically plausible neural network and required iterations in the previous similarity matching networks. However, Theorems \ref{thm:psp1} and \ref{thm:psw1} show that at fixed points $\m{M}\fp$ is diagonal, which forms the basis of our approach for iteration-free dynamics.  Formally, we split $\m{M}(s)$ into its diagonal and off-diagonal parts as
\begin{equation}
  \m{M}(s) = \m{M}_\text{d}(s) + \m{M}_\text{o}(s),
\end{equation}
where $\m{M}_\text{d}(s)$ is diagonal and $\m{M}_\text{o}(s)$ is zero on the diagonal.  Assuming that $\m{M}(s)$ is not far from diagonal at any point in time, we observe from the first-order Taylor series approximation of the matrix inverse that
\begin{equation}\label{eq:taylor}
  \m{M}(s)^{-1} \approx \m{M}_\text{d}(s)^{-1} - \m{M}_\text{d}(s)^{-1}\m{M}_\text{o}(s)\m{M}_\text{d}(s)^{-1}
\end{equation}
with approximation error $\O(\|\m{M}_\text{o}(s)\|^2)$.  While this approximation on the surface still involves matrix inversion, the only matrix to be inverted is $\m{M}_\text{d}$, which is diagonal and thus may be inverted via element-wise inversion of the diagonal.  Because no {\it bona fide} matrix inversion is required, fast fixed-point iterations are no longer necessary, and so we refer to this as an ``iteration-free'' approximation.

Using \eqref{eq:taylor} we obtain the modified dynamical system for PSP
\begin{equation}\label{eq:dyn2psp}
  \begin{array}{r@{}l}
  \m{Y}(s) &{}= \left[\m{I} - \m{M}_\text{d}^{-1}(s)\m{M}_\text{o}(s)\right]\m{M}_\text{d}^{-1}(s)\m{W}(s)\m{X},\\[3pt]
  \frac{d\m{W}(s)}{ds} &{}= \m{Y}(s)\m{X}^\top - \m{W}(s),\\[3pt]
  \tau\frac{d\m{M}(s)}{ds} &{}= \m{Y}(s)\m{Y}(s)^\top - \m{\Lambda}\m{M}(s)\m{\Lambda},
\end{array}
\end{equation}
which arises from inserting \eqref{eq:taylor} into \eqref{eq:dyn1psp}.

\begin{thm}\label{thm:psp2}
  Suppose the same setting as Lemma \ref{lem:uniquepsp}.  Then any stable fixed point $(\m{M}^*,\m{W}^*)$ of the dynamical system \eqref{eq:dyn1psp} is also a stable fixed point of \eqref{eq:dyn2psp}.
\end{thm}

\begin{proofatend}
  The linearization of \eqref{eq:dyn2psp} agrees with that of \eqref{eq:dyn1psp} around any fixed point with diagonal $\m{M}$.
\end{proofatend}

Following the same procedure we obtain a modified dynamical system for PSW,
\begin{equation}\label{eq:dyn2psw}
  \begin{array}{r@{}l}
  \m{Y}(s) &{}= \left[\m{I} - \m{M}_\text{d}^{-1}(s)\m{M}_\text{o}(s)\right]\m{M}_\text{d}^{-1}(s)\m{W}(s)\m{X},\\[3pt]
  \frac{d\m{W}(s)}{ds} &{}= \m{Y}(s)\m{X}^\top - \m{W}(s),\\[3pt]
  \tau\frac{d\m{M}(s)}{ds} &{}= \m{Y}(s)\m{Y}(s)^\top - \m{\Lambda}^2,
\end{array}
\end{equation}
for which we can prove a stronger result.

\begin{thm}\label{thm:psw2}
  Suppose the same setting as Lemma \ref{lem:uniquepsp}.  Then the stable fixed points $(\m{M}^*,\m{W}^*)$ of the dynamical system \eqref{eq:dyn2psw} are exactly the stable fixed points of \eqref{eq:dyn1psw}.
\end{thm}

\begin{proofatend}
  The fixed point conditions for $\eqref{eq:dyn2psw}$ imply $\m{M}_\text{d}^{-1} - \m{M}_\text{d}^{-1}\m{M}_\text{o}\m{M}_\text{d}^{-1}$ is diagonal at a fixed point. Using this, we see that the linearization of \eqref{eq:dyn2psw} around any fixed point is identical to the linearization of \eqref{eq:dyn1psw}.
\end{proofatend}

\subsection{Online Algorithms}
The dynamical system \eqref{eq:dyn2psp} suggests an online algorithm for PSP by replacing the prescribed gradient dynamics for $\m{M}$ and $\m{W}$ with stochastic gradient updates.  In Algorithm \ref{alg:psp}, we outline the computations required at the presentation of each input vector $\m{x}_t$ to compute the corresponding output $\m{y}_t$ and the updates to the matrices $\m{M}$ and $\m{W}$ with asymptotic complexity $\O(KN)$ per iteration.  We note that in the online algorithm we have rescaled $\m{M}$ and $\m{W}$ by a factor of $T$ for numerical convenience.
\begin{algorithm}[H]
 \caption{Biologically Plausible Online PSP}
 \begin{algorithmic}\label{alg:psp}
 \renewcommand{\algorithmicrequire}{\textbf{Input:}}
 \REQUIRE Initial weights $\m{M}\in\R^{K\times K}$ and $\m{W}\in\R^{K\times N}$, diagonal $\m{\Lambda}\in\R^{K\times K}$
  \FOR {$t = 1,2,3,\dots$}
  \STATE\texttt{// Neural dynamics (two-step)}
  \STATE $\v{\tilde y}_t \gets \m{M}_\text{d}^{-1}\m{W} \v{x}_t$
  \STATE $\v{y}_t \gets \m{M}_\text{d}^{-1}\m{W} \v{x}_t - \m{M}_\text{d}^{-1}\m{M}_\text{o}\v{\tilde y}_t$
  \STATE\texttt{// Synaptic plasticity}
  \STATE $\m{W} \gets \m{W} + \alpha_t\left(\v{y}_t\v{x}_t^\top - \m{W}\right)$
  \STATE $\m{M} \gets\m{M} + \tau^{-1}\alpha_t\left(\v{y}_t\v{y}_t^\top - \m{\Lambda}\m{M}\m{\Lambda}\right) $
  \ENDFOR
 \end{algorithmic}
 \end{algorithm}
Similarly, from the modified PSW dynamical system \eqref{eq:dyn2psw} we obtain Algorithm \ref{alg:psw}, again with per-iteration complexity $\O(KN)$.

\begin{algorithm}[H]
 \caption{Biologically Plausible Online PSW}
 \begin{algorithmic}\label{alg:psw}
 \renewcommand{\algorithmicrequire}{\textbf{Input:}}
 \REQUIRE Initial weights $\m{M}\in\R^{K\times K}$ and $\m{W}\in\R^{K\times N}$, diagonal $\m{\Lambda}\in\R^{K\times K}$
  \FOR {$t = 1,2,3,\dots$}
  \STATE\texttt{// Neural dynamics (two-step)}
  \STATE $\v{\tilde y}_t \gets \m{M}_\text{d}^{-1}\m{W} \v{x}_t$
  \STATE $\v{y}_t \gets \m{M}_\text{d}^{-1}\m{W} \v{x}_t - \m{M}_\text{d}^{-1}\m{M}_\text{o}\v{\tilde y}_t$
  \STATE\texttt{// Synaptic plasticity}
  \STATE $\m{W} \gets \m{W} + \alpha_t\left(\v{y}_t\v{x}_t^\top - \m{W}\right)$
  \STATE $\m{M} \gets\m{M} + \tau^{-1}\alpha_t\left(\v{y}_t\v{y}_t^\top - \m{\Lambda}^2\right) $
  \ENDFOR
 \end{algorithmic}
 \end{algorithm}

 While the online algorithms fundamentally stem from applying stochastic dynamics to a saddle point problem (and thus convergence is not assured), we show in Section \ref{sec:exp} that this is not an issue in practice.

\subsection{Biologically Plausible Implementation}
\begin{figure}
  \centering
  \includegraphics[scale=0.4]{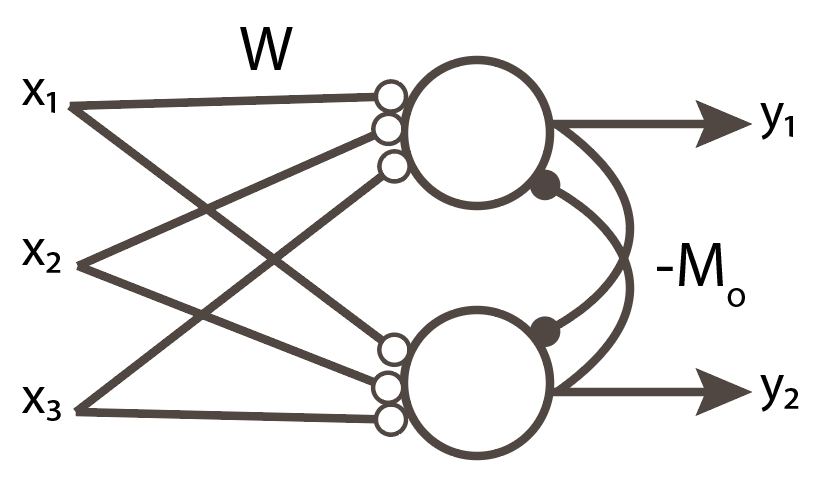}
  \caption{\label{fig:top} The topology for a biological implementation of Algorithms \ref{alg:psp} and \ref{alg:psw} with three-dimensional input vector $\v{x}\in\R^3$ and two-dimensional output vector $\v{y}\in\R^2$ (here, the dependence on $t$ is omitted)}
\end{figure}

Algorithms \ref{alg:psp} and \ref{alg:psw} can be naturally implemented by a neural network of the topology shown in Figure \ref{fig:top}. At the presentation of each  $\v{x}_t$, the network operates in two phases. First, the output vector $\v{y}_t$ is computed by multiplying the input $\v{x}_t$ by the matrix $\m{W}$ encoded in feed-forward synaptic weights. The diagonal elements of the matrix $\m{M}_\text{d}$ (not shown in figure) are interpreted as a single scaling variable per neuron equalizing the output activity level of that neuron. The ``off-diagonal matrix'' $\m{M}_\text{o}$ maps to lateral weights between distinct neurons $\v{y}_i$ and $\v{y}_j$, corresponding to inhibitory connections between neurons in the same layer. In the second phase, $\m{M}$ and $\m{W}$ are updated using Hebbian rules to reflect the new input-output pair.

While the network topology in Figure \ref{fig:top} is shared with the original PSP and PSW implementations \cite{derivation,whydo} and is standard in the computational neuroscience literature \cite[Section 8.3]{dayan}, the two-step dynamics described in Algorithms \ref{alg:psp} and \ref{alg:psw} imply a non-standard model for the neural activity (though other two-step dynamics with a similar flavor have appeared as, e.g., the learning/unlearning phases of Linsker \cite{linsker2005improved}).  This stems directly from the lack of recurrent dynamics.  Considering the neural dynamics in either algorithm,  the two-step dynamics can be viewed as a form of time multiplexing on the topology of Figure \ref{fig:top}, where the contribution from the feed-forward input is computed first to construct the initial signal $\v{\tilde y}_t$, then this initial signal is propagated laterally and combined with the feed-forward input to produce the output signal $\v{y}_t$.

Importantly, the synaptic plasticity rules in both algorithms are entirely local with respect to the proposed architecture.

\section{Numerical Experiments}\label{sec:exp}
To demonstrate the effectiveness of our proposed algorithms numerically, we apply them to a few synthetic data examples.


As a measure of convergence we consider error metrics which we term the subspace alignment error:
$$
E_\text{Pro}\left(\m{\widehat U}_K\right) \equiv \min_{\m{Q}\in\mathbb{O}_N}\frac{\left\|\m{\widehat U}_K\m{Q} - \m{U}_K \right\|^2}{\left\|\m{U}_K\right\|^2},
$$
where $\mathbb{O}_N$ is the set of $N\times N$ orthogonal matrices, the optimal $\m{Q}$ rotates $\m{\widehat U}_K$ to best align with $\m{U}_K$ by solving the orthogonal Procrustes problem \cite{pro}, and the appropriate definition of $\m{\widehat U}_K$ in terms of the matrices $\m{M}$ and $\m{W}$ depends on the algorithm being evaluated (see Lemmas \ref{lem:uniquepsp} and \ref{lem:uniquepsw}).

We investigate the performance of versions of the proposed algorithms both with and without the Taylor series approximation in \eqref{eq:taylor}, leading to the following four online algorithms:

\begin{quote}
 {\bf ifPSP} The online algorithm for iteration-free PSP as described in Algorithm \ref{alg:psp}, with
  $$
  \m{\widehat U}_K^\top \equiv \m{\Lambda}^{-1}(\m{M}_\text{d}^{-1} - \m{M}_\text{d}^{-1}\m{M}_\text{o}\m{M}_\text{d}^{-1})\m{W}.
  $$

 {\bf PSP}  An online algorithm for PSP with general matrix inversion (which would require iteration), obtained by replacing the neural dynamics in Algorithm \ref{alg:psp} with $\v{y}_t~\hspace{-0.3cm}\gets\hspace{-0.3cm}~\m{M}^{-1}\m{W}\v{x}_t$.  Here we take $
    \m{\widehat U}_K^\top \equiv \m{\Lambda}^{-1}\m{M}^{-1}\m{W}$.

    \vspace{0.2cm}

   {\bf ifPSW} The online algorithm for iteration-free PSW as described in Algorithm \ref{alg:psw}, with
  $$
  \m{\widehat U}_K^\top \equiv \m{\Sigma}_K\m{\Lambda}^{-1}(\m{M}_\text{d}^{-1} - \m{M}_\text{d}^{-1}\m{M}_\text{o}\m{M}_\text{d}^{-1})\m{W}.
  $$

   {\bf PSW} An online algorithm for PSW with general matrix inversion (which would require iteration), obtained by replacing the neural dynamics in Algorithm \ref{alg:psw} with $\v{y}_t\hspace{-0.1cm}\gets\hspace{-0.1cm}\m{M}^{-1}\m{W}\v{x}_t$.  Here we take $
    \m{\widehat U}_K^\top~\equiv~\m{\Sigma}_K\m{\Lambda}^{-1}\m{M}^{-1}\m{W}$.
\end{quote}

For comparison we also look at the performance of ``offline'' versions of the above algorithms.  For example, to construct an offline version of Algorithm \ref{alg:psp}, suppose $\mathbb{E}\left[\v{x}_t\v{x}_t^\top\right]=\m{G}$.  Then we take the synaptic plasticity rules and replace $\v{y}_t\v{x}_t^\top$ and $\v{y}_t\v{y}_t^\top$ with their expectations over the data
\begin{align*}
\mathbb{E}\left[\v{y}_t\v{x}_t^\top\right] &= \m{F}\m{G} \;\; \text{and} \;\; \mathbb{E}\left[\v{y}_t\v{y}_t^\top\right] = \m{F}\m{G}\m{F}^\top,
\end{align*}
where $\m{F}=\left[\m{I} - \m{M}_\text{d}^{-1}\m{M}_\text{o}\right]\m{M}_\text{d}^{-1}\m{W}$ is the neural filter mapping inputs to outputs at the current iteration (i.e., $\m{F}$ is such that in the online algorithm  $\v{y}_t = \m{F}\m{x}_t$).
This leads to four offline algorithms, one for each of the four online algorithms above.

\subsection{Parameters and Initialization}
For each example we sample the data $\{\v{x}_t\}_{t=1}^T$ independently from a multivariate Gaussian distribution $\v{x}_t~\sim~\mathcal{N}(\v{0},\m{G})$ with specified population covariance matrix $$\m{G}=\m{R}\m{\widetilde G}\m{R}^\top,$$ where $\m{R}\in\mathbb{O}_N$ is a random orthogonal matrix (Haar measure) and $\m{\widetilde G}$ is diagonal.

We consider two different problem sizes.  For the ``smaller' problem, we set the input dimension $N=10$ and output dimension $K=3$ and choose $\m{\widetilde G}_{11} = 1$, $\m{\widetilde G}_{22}=0.75$, $\m{\widetilde G}_{33}=0.5$, and $\m{\widetilde G}_{kk} = 0.2$ otherwise.  We use $\m{\Lambda}=\text{diag}([1,\,0.85,\,0.7])$ and  $\alpha_t = 10/(250 + t)$ for the small online tests, choosing $\tau=0.5$ for PSP-based algorithms and $\tau=1$ for PSW-based algorithms.  For the ``larger'' problem, we set the input and output dimensions as $N=100$ and $K=10$.  The singular values of the population covariance matrix of $\m{X}$ are given by
 $$
 \m{\widetilde G}_{kk} = \left\{\begin{array}{cc}
 1-\frac{k-1}{2(K-1)}&k\le 10, \\
 0.02& \text{else.}\end{array} \right.
 $$
 We use $\m{\Lambda}_{kk}=1-\frac{3(k-1)}{10(K-1)}$ and choose the step sequence
 $$
 \alpha_t = \left\{\begin{array}{cc}
 1.1\times 10^{-3}& t \le 10000,\\
 1.0\times 10^{-4} & \text{else,}\end{array} \right.
 $$for PSP-based algorithms and $\alpha_t=1.0\times10^{-3}$ for PSW-based algorithms.  We keep $\tau$ as in the small problem.

 Because the offline algorithms are not stochastic, the step sequence can be taken much larger than in the online case.  Thus, we use the step rule $\alpha_t = 1.0\times10^{-1}$ for both the smaller and larger problem in the offline setting, which gives convergence to the principal subspace in a relatively small number of iterations, $T$.  Note that, because the offline algorithms do not sample the data but rather use the covariance matrix directly, $T$ does not correspond to number of samples used but rather only to iteration count.

 For initialization in both the smaller and larger problems, we use $\m{M} = \m{I}$ for the PSP-based algorithms and $\m{M}=0.3\m{I}$ for the PSW-based algorithms.  In all cases, $\m{W}$ is initialized to have random normal entries with mean zero and variance $1/N$.

\subsection{Results}

\begin{figure}
  \centering
  \includegraphics[scale=0.37]{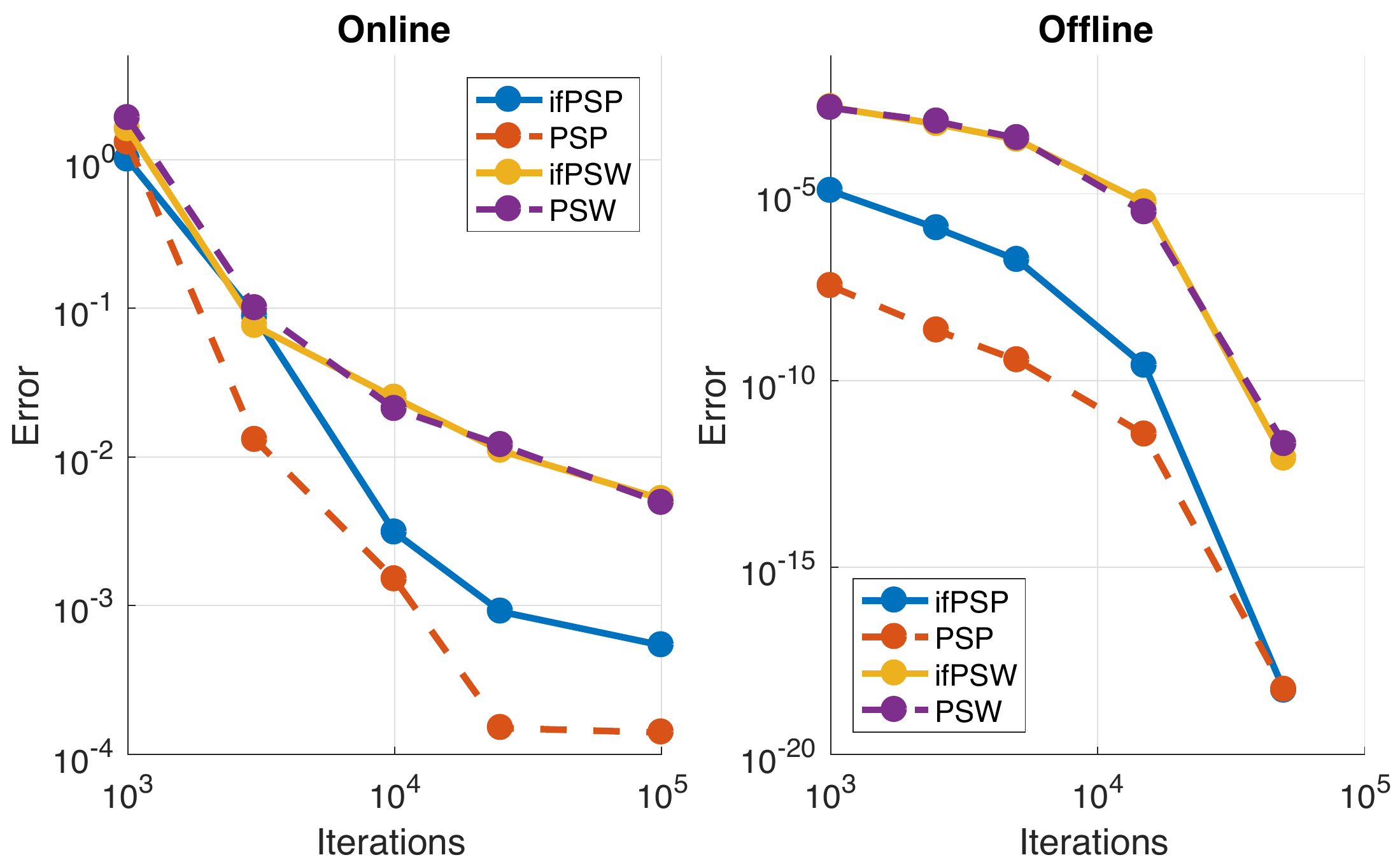}
    \caption{\label{fig:plt} On the left, we show the online results for the larger problem ($N~=~100$) as in Table \ref{tab:resultson}.  On the right, we show the offline results for the larger problem as in Table \ref{tab:resultsoff}.  In all cases, the error is the appropriate Procrustes error $E_\text{Pro}$.}
\end{figure}

In Table \ref{tab:resultson}, we show the subspace alignment error, $E_\text{Pro}$ for the four different online algorithms for varying numbers of points, where each error is computed as the median across 100 trials (see also Figure \ref{fig:plt}, left).  For both the smaller ($N=10$) and larger ($N=100$) problem, we observe gradual convergence of the online algorithms, with the new iteration-free algorithms ifPSP and ifPSW behaving similarly to the variants PSP and PSW using full general matrix inversion at each iteration.

\begin{table}
  \caption{The subspace alignment error $E_\text{Pro}$ for the online algorithms on examples of two different sizes\label{tab:resultson}}
  \setlength{\tabcolsep}{0.25em}
\begin{tabular}{@{}lcccc@{\hskip 0.15in}cccc@{}}
\toprule
&\multicolumn{4}{c}{$N=10,K=3$}&\multicolumn{4}{c}{$N=100,K=10$}\\
               & ifPSP    & PSP & ifPSW & PSW & ifPSP & PSP & ifPSW & PSW  \\ \midrule
T=1000         &  2.1e-2& 1.9e-2 & 9.6e-1 & 7.7e-1  &  1.0e-0& 1.3e-0  &1.6e-0  &1.9e-0  \\
T=10000        & 1.5e-4 &   4.1e-4& 1.3e-2  & 1.6e-2  & 3.1e-3& 1.5e-3  & 2.5e-2 &2.1e-2   \\
T=100000        & 1.7e-5 &   5.5e-5& 1.8e-3  & 1.8e-3  & 5.4e-4& 1.4e-4  & 5.2e-3 &4.9e-3   \\ \bottomrule
\end{tabular}
\end{table}

To validate our theoretical results in the offline setting, we give corresponding results of the error $E_\text{Pro}$ for the \emph{offline} algorithms in Table \ref{tab:resultsoff} (see also Figure \ref{fig:plt}, right).  As in the online case, we see that ifPSP and ifPSW behave similarly to PSP and PSW (respectively), though now in the offline case we observe fast convergence to high precision in a relatively small number of iterations due to the lack of stochasticity.

Overall, we see that there is no marked increase in our error measure when using ifPSP or ifPSW when compared to PSP or PSW and we observe convergence to the principal subspace in all cases.  The offline results show fast convergence of the dynamical system to a stable fixed point.  In the online case, convergence is naturally slower due to the variance inherent in stochastic dynamics (and correspondly small step size), but still evident.

\begin{table}
  \caption{The subspace alignment error $E_\text{Pro}$ for the offline algorithms on examples of two different sizes\label{tab:resultsoff}.  The entries denoted by $*$ are computed to be less than $10^{-18}$.}
    \setlength{\tabcolsep}{0.25em}
\begin{tabular}{@{}lcccc@{\hskip 0.15in}cccc@{}}

\toprule
&\multicolumn{4}{c}{$N=10,K=3$}&\multicolumn{4}{c}{$N=100,K=10$}\\
               & ifPSP & PSP & ifPSW & PSW & ifPSP & PSP & ifPSW & PSW  \\ \midrule
               T=100          &2.7e-5  & 2.3e-4 &9.5e-3  & 9.8e-3  & 6.0e-4 & 5.3e-6  &1.3e-2&1.4e-2 \\
               T=1000         &  5.9e-10& 2.3e-10 & 4.2e-7 & 5.5e-7  &  1.2e-5& 3.4e-8  &2.1e-3  &2.0e-3  \\
               T=5000      & * &  * & *  & *  & 1.7e-7& 3.5e-10  & 2.8e-4 &3.1e-4 \\
               T=50000     & * & * & * & *  & *  &  * & 8.2e-13  &2.0e-12  \\ \bottomrule \end{tabular}
\end{table}

\section{Discussion}
In this work we proposed a modification of the similarity matching framework for online PCA to develop online PCA algorithms that do not require general matrix inversion with the arrival of each new input vector.  Following the same mechanics as the original framework, we proved rigorous stability results for dynamical systems stemming from the modified objectives, showing that gradient dynamics on a saddle point formulation of a modified similarity matching objective has stable fixed points corresponding to the principal subspace.  The resulting algorithms closely resemble the original ones, but, importantly, do not require iterations to invert a general (non-diagonal) matrix and thus have greater biological plausibility.

Our numerical results show the inversion-free algorithms presented here give similar performance in terms of error to corresponding variants using matrix inversion.  However, in contrast to the algorithms based on inversion, the per-iteration computational complexity of the new algorithms is only $\O(KN)$ versus $\O(KN + K^3)$, which is lower in certain regimes.

Unlike previous similarity-based networks the iteration-free algorithm does not suffer from the degeneracy of the output with respect to a multiplication by an orthogonal matrix. Uniqueness of the solution potentially simplifies the analysis downstream.  Further, while the proposed algorithms are not expected to obtain true ``convergence'' except perhaps in the case of infinite stationary data, it is interesting to observe that the prediction of stable fixed points with $\m{M}_\text{o}=\m{0}$ implies a pruning of the network topology in Figure \ref{fig:top}.  In other words, at convergence, the lateral connections should disappear and near fixed points the lateral synaptic weights should be small.

Because the proposed online algorithms are based on simultaneous gradient ascent-descent dynamics on a min-max problem, it is difficult to prove convergence results for the online algorithms presented here.  In future work, we hope to use results like those used for Generative Adversarial Networks (GANs) \cite{ttur} to develop a theory of convergence for the online case, whether for these algorithms or for modified versions as appropriate.

\section*{Acknowledgments}
The authors thank Mariano Tepper and Anirvan Sengupta for useful discussion that contributed to the quality of this manuscript.

\bibliographystyle{IEEEtran}
\bibliography{pca}

\appendix

\section{Proofs}
\printproofs

\end{document}